# How the Natural Interpretation of QM Avoids the Recent No-go Theorem

Anthony Rizzi[1]

[1]*Institute for Advanced Physics, Baton Rouge, LA 70895*

**Abstract:** A recent no-go theorem gives an extension of the Wigner's friend argument that purports to prove that "Quantum theory cannot consistently describe the use of itself." The argument is complex and thought provoking, but fails in a straightforward way if one treats QM as a statistical theory in the most fundamental sense, i.e. if one applies the so-called ensemble interpretation. This explanation is given here at an undergraduate level, which can be edifying for experts and students alike. A recent paper has already shown that the no-go theorem is incorrect with regard to the de Broglie Bohm theory and misguided in some of its general claims. This paper's contribution is three fold. It shows how the extended Wigner's friend argument fails in the ensemble interpretation. It also makes more evident how natural a consistent statistical treatment of the wave function is. In this way, the refutation of the argument is useful for bringing out the core statistical nature of QM. It, in addition, manifests the unnecessary complications and problems introduced by the collapse mechanism that is part of the Copenhagen interpretation. The paper uses the straightforwardness of the ensemble interpretation to make the no-go argument and its refutation more accessible.

## Introduction

The recent no-go theorem by D. Frauchiger and R. Renner (DF&RR) [1] makes the strong claim that quantum mechanics is violated in a certain case when multiple observers are making predictions. These claims have been successfully challenged by D. Lazarovici and M. Hubert. [2] These authors point out that the no-go arguments do not apply to the de Broglie Bohm interpretation because it has no collapse mechanism and thus does not fall under the implicit assumptions of the proof. Here we show that the theorem fails more generally since its implicit assumptions do not apply to the ensemble interpretation, which is the natural statistical understanding of QM (therefore, in principle, allows no collapse).[3] The ensemble interpretation is a newer approach first championed generically by Einstein but only developed formally by L. Ballentine,[4] starting with his seminal paper in 1970.[5] This approach has been, recently, more specifically developed in a new textbook.[6,7] The ensemble approach simply treats the wave function as a device for describing the quantum state statistically. The only evolution of the wave function is that due to the Schrödinger equation. A given measurement establishes a one-to-one relationship between 1) the ontic value of a certain observable of, e.g., a specific particle of an ensemble whose statistical state is described by the wave function and 2) the ontic value of the "pointer" of a member of the ensemble of measuring devices. Because the measurement requires an interaction, the wave functions describing the ensembles of measuring devices and the particle become entangled during the measurement. The evolution of the state of the particle/device system before, during and after the measurement is described solely by the Schrödinger equation (SE). Thus, it does not, in general, collapse to the value being measured because the SE does not in general predict this.

---

[1] D. Frauchiger, R. Renner, *Quantum theory cannot consistently describe the use of itself*, Nat. Commun. **9** No. 1038 (2018).

[2] D. Lazarovici, M. Hubert, *How Quantum Mechanics can consistently describe the use of itself*, Sci. Rep. **9** No. 470, (2019)

[3] A. Rizzi, *A Simple Approach To Measurement in Quantum Mechanics*, to be published.

[4] L. E. Ballentine, *Quantum Mechanics: A Modern Development* (World Scientific Publishing, Singapore,1998)

[5] L. E. Ballentine, *The Statistical Interpretation of Quantum Mechanics*, Rev. Mod. Phys. 42 No. 4, 358-381 (1970)

[6] A. Rizzi, *Physics for Realists: Quantum Mechanics* (IAP Press, Baton Rouge, 2018).

[7] The ensemble interpretation is used to explain the real meaning PBR theorem in A. Rizzi, *Does the PBR Theorem Rule out a Statistical Understanding of QM?* Found. of Phys., **48** (12), 1770-1793.

The DF&RR no-go theorem is an extended Wigner's friend argument. It is, thus, helpful to review the original (un-extended) Wigner friend argument.[8] In it, Wigner argued as follows for a consciousness-driven collapse of the wave function during a measurement. Suppose one is given a particle in a spin right state, so that we have $|\psi_1\rangle = |\rightarrow\rangle$. Wigner's friend, $F$, using an up/down Stern Gerlach (SG) device will either measure it to be up or down. Then, according to the "collapse to an eigenstate on measurement" rule: after the measurement, the *friend*, $F$, will say the system state is either:

(1) $\qquad\qquad |\uparrow\rangle|\psi_\uparrow\rangle|light\rangle|F_\uparrow\rangle \quad or \quad |\downarrow\rangle|\psi_\downarrow\rangle|no-light\rangle|F_\downarrow\rangle.$

Here $F_\uparrow$ ($F_\downarrow$) represent the brain states of $F$ and light (no-light) represent the state of the light used to detect the particle in the up (down) part of the SG.

However, *Wigner*, who knows $F$ did the experiment but not the results, calculates the state to be:

(2) $\qquad\qquad \frac{1}{\sqrt{2}}\left(|\uparrow\rangle|\psi_\uparrow\rangle|light\rangle|F\rangle_\uparrow + |\downarrow\rangle|\psi_\downarrow\rangle|no-light\rangle|F\rangle_\downarrow\right)$

Without something to distinguish the states of the two observers, they must be the same and thus the system cannot be represented by both (1) and (2). One can resolve the contradiction by invoking consciousness collapse, i.e., the idea that $F$ collapses the state because he knows the value of the measurement, but $W$ does not. Notice how, in this ad hoc resolution, each person, in a way, creates his own separate world![9]

Now, the extended Wigner's friend argument involves four experimenters. There are two labs with one experimenter in each lab, and one supervisor for each lab. The experimenters are given names and diagrammed in Figure 1. The extended Friend argument shows: 1) $W$ 's calculations will directly predict that he will measure his lab (his friend, $F$, and the things $F$ measures) to be in a state called "*ok*," 2) while $\overline{F}$ will predict that $W$ will measure the opposite state (called *fail*) and, indeed, using $\overline{F}$, $F$ and $\overline{W}$ 's predictions, $W$ would predict himself to measure fail. In short, proceeding directly from his own perspective, $W$ predicts the state of his lab to be ok, whereas proceeding through $\overline{F}$, $F$ and $\overline{W}$ 's perspectives he predicts he will see fail. This is a clear contradiction.

---

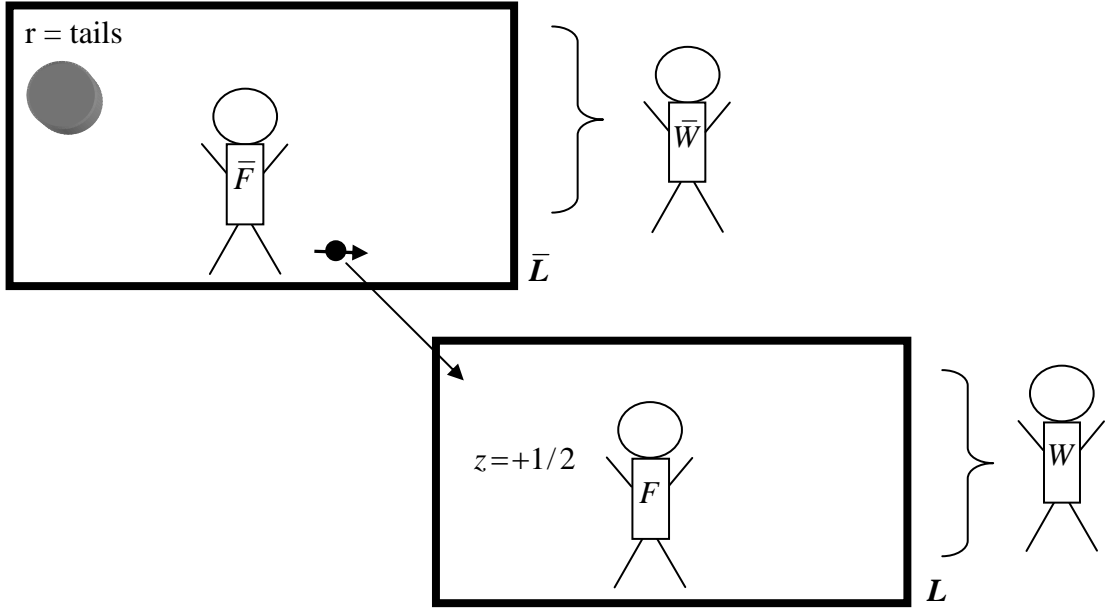

**Figure 1**: Wigner's brother, $\bar{W}$, supervises a lab ($\bar{L}$) in which a friend, $\bar{F}$, does an experiment. Wigner himself, *W*, supervises a different lab (*L*) in which his own friend, *F*, does an experiment.

The core of this paper begins in the next section with a top level description of the extended Wigner's friend thought experiment. That top-level description gives the argument, which we summarized above, that purportedly shows that different observers, each properly applying quantum theory, can yield results that contradict each other. It is as if observer *A* applies Newtonian mechanics to a body and predicts that he will see the body accelerate *upward* while observer *B* predicts that observer *A* will see the body accelerates *downward*.

The arguments in the top level view section use the calculations given in the sections that follow it. In particular, just after the top level argument, there is a section that gives the details of the evolution of the system through the various steps of the thought experiment for the case of the ensemble approach in which there is no collapsing of the wave function, but only pure Schrödinger equation evolution. The next section gives the evolution of the system through the various steps of the thought experiment for the case when $\bar{F}$ collapses the state of the coin to tails, which is needed for some of the perspectives used in the DF&RR approach. In these last two sections, equations are given for *each* step of the experiment, even those ignored in DF&RR's formal analysis; this is done especially in order manifest the nature of the ensemble approach in which *every* measurement entangles the measuring device and the thing measured. A final section summarizes the analysis.

### The Extended Wigner's Friend (EWF) Argument, Top View

To make the argument of the no-go paper[1] more accessible, I break the argument into blocks, drawing out the key points of the argument and where it fails for the ensemble interpretation. Note that DF&RR make implicit assumptions about the nature of measurement. In what follows, I manifest the standard textbook assumptions that follow their reasoning and yield their result.

The experiment proceeds as follows. Wigner, *W*, oversees a lab, *L*, and his twin brother, $\bar{W}$, oversees his own lab, $\bar{L}$, pictured in Figure 1. Notice Wigner (*W*) has a friend, *F*, that runs his lab and his brother ($\bar{W}$) has a friend, $\bar{F}$ that runs his.

A series of instructions is given to the four participants. Everyone knows what those instructions are. They are (see box 1 DF&RR) as follows. The variables and wave functions referred to below are defined and calculated, respectively, in later sections.

### EWF Instructions:

At time 00, Agent $\bar{F}$ will measure coin and then set the spin of a particle to spin right ($\rightarrow$) if it comes up tails, and to spin down ($\downarrow$) if its heads. This gives for the overall state, in the ensemble interpretation: $\psi_2$.

At time 10, agent *F* will measure the spin S, giving $\psi_3$ for the overall system state; *F* measures spin up.

At time 20, Agent $\bar{W}$ measures whether his barred lab ($\bar{L}$) is ok or not (which we call, respectively $\overline{ok}$ or $\bar{f}$). $\overline{ok}$ is a particular state of $\bar{F}$'s brain *and* the coin in the barred lab, giving $\psi_4$. After this, $\bar{W}$ announces his result to *W*, giving $\psi_5$.

At time 30, *W* measures whether his unbarred lab (*L*) is ok or not (which we call *ok* or *f*), giving $\psi_6$, then he announces his result to $\bar{F}$, giving $\psi_7$.

At time 40, if $\bar{W}$ measures $\overline{ok}$ and *W* measures *ok* then the experiment stops if not one cycles back to time 00 and repeats the steps.

In what follows we consider only that pass through the above sequence that reaches the halting condition (i.e., $\bar{W}$ measures $\overline{ok}$ **and** *W* measures *ok*).

To start, we analyze what *W* will see from two points of view: from $\bar{F}$'s perspective or from that of *W* himself. Then, we bring in *F* and $\bar{W}$'s perspective and generate our direct contradiction.

### A. From $\bar{F}$'s perspective:

$\bar{F}$ *flips* his quantum coin and gets tails. DF&RR assume (as Copenhagen Interpretation does) that $\bar{F}$'s knowledge collapses the wave function (for him) to $\psi_{c2}$ (see section below titled **EWF State Predictions II**) --where the subscript *c* indicates that we are speaking about a wave function that assumes this collapse. We can then predict that, when *W* measures his lab with an instrument calibrated to readout either *ok* or fail (*f*), the instrument will read "fail".[10] We can see this by noting that $\langle \psi_{c7} | ok \rangle_L = 0$. (see equation (21) in **EWF II**). Since the *ok/f* is a complete orthonormal basis, *W* will read fail.

**B. From *W*'s perspective:**

    *W* knows that $\bar{W}$ reads the $\overline{ok}$ state because it is announced to him at time 20. And, since *W* has not observed the state of the coin in $\bar{L}$, it seems that the wave function at time 10 is therefore not collapsed, i.e. not forced into an eigenstate in the heads/tails basis (for he does not know the result of this state). Thus, the wave function is $\psi_2$, as given in the section titled **EWF I** and ***not*** $\psi_{c2}$. This different starting wave function, in turn, means the state of the system is also different for each later step of the experiment listed above. Those wave functions are also given in section **EWF I** and are labeled *without* the subscript c, as no collapse is assumed. Now, using these different states, one can see that *W* will predict that at time 40, there is a 1/12 chance that *W* will measure *ok*,[11] (that state of *W* knowing the lab (*L*) to be *ok*, we label $W_{ok}$) and $\bar{W}$ will measure $\overline{ok}$ (which we label $\bar{W}_{\overline{ok}}$). We see this by noting that: $\langle \psi_7 | (|W_{ok}\rangle |\bar{W}_{\overline{ok}}\rangle) \rangle = 1/\sqrt{12}$ --see equation (12), (to avoid notational clutter, we have omitted the $ok/\overline{ok}$ states and some of the ket states associated with *W* and $\bar{W}$ --which have components which attach to *ok* and $\overline{ok}$ respectively). Thus, it is indeed possible to meet the halting condition, so we assume we have. Hence, **W predicts he will measure his lab to be in the *ok* state**. However, as seen in the last line in subsection A above, from $\bar{F}$'s perspective *W* will read fail! We already have a contradiction of sorts, but you may argue they are in different worlds. So, following DF&RR, let us bring it home.

**C. Bringing in *F* and $\bar{W}$ 's perspective**,

    Assuming *F* measures the spin to be up, then, by $\psi_3$ (*F* does not measure the coin directly so apparently doesn't collapse it), *F* knows immediately after time 10 that $\bar{F}$ knows that the coin toss gave tails, so *F* can deduce, using the "collapsed" wave function, $\psi_{c7}$ (note location of $|\bar{F}W_f\rangle$ in equation (21)), that $\bar{F}$ will think (right after time 30) that *W* will measure *L* to be in the fail state, so *F* himself obviously knows *W* will measure *L* to be in the fail state as well.

    Assuming, as we posited above, that $\bar{W}$ measures $\bar{L}$ to be in the $\overline{ok}$ state then $\bar{W}$ can deduce that from $\psi_4$ ($\bar{W}$ does not collapse the spin or the coin measurement) that *F* knows that the spin of the particle is up, from which $\bar{W}$ can deduce, following *F*'s reasoning in previous paragraph, that *F* will think that *W* measures *L* to be in the fail state. And, thus, we have that $\bar{W}$ thinks *W* will measure *L* in the fail state.

---

[10] In our analysis, we do not formally breakout *W*'s measurement instrument. One can think of the *W* in our equations as representing the physical state of *W*'s brain (including all the relevant neurological variables) along with the state of the instrument he uses.

[11] Note, one gets 1/12 whether one allows *W* to collapse his measured systems or not.

Lastly, because $\bar{W}$ announces to $W$ that he has found his lab's state to be in $\overline{ok}$, $W$ can deduce that $\bar{W}$ will think he ($W$ himself) will observe his lab to be in the fail state. **So, $W$ himself thinks his lab is in the *fail* state after time 30. But, arguing directly from $W$'s perspective in B above, we found $W$ thought his lab was in the *ok* state! A direct contradiction! Logically, $W$ may have his own world but it better be consistent with itself!**

You see the collapse mechanism has made a mess of the analysis. Once we recognize that there is no collapse, this mess goes away and everything falls out. We only need to use the states represented by the wave function in the non-collapse (ensemble interpretation) section, **EWF I**. In this case, the arguments in subsection A do not go through and we see that $\langle \psi_7 | W_{ok_L} \rangle \neq 0$, so $W$ can measure his lab to be in ok or fail! The reasoning in subsection B is unchanged except for the conclusion based on subsection A. In subsection C, the reasoning of the first and second paragraphs are incorrect because they are based on the collapsed wave function and reasoning from subsection A. The net result is that again we *cannot* conclude that $W$ will necessarily measure his lab to be in the fail state, so there is no longer any contradiction.

Now, having seen the argument and its resolution in top view, we move to the details of the calculation of the state evolution, first in the case of the simple Schrödinger equation evolution of the ensemble interpretation and then in the case of collapse.

## The EWF State Predictions I
## (no collapse, ensemble interpretation)

The experiment proceeds as follows. Recall Wigner, $W$, oversees a lab, $L$, and his twin brother, $\bar{W}$, oversees his own lab, $\bar{L}$, pictured in Figure 1. Wigner ($W$) has a friend, $F$, that runs his lab and his brother ($\bar{W}$) has a friend, $\bar{F}$ that runs his.

We start in the top lab ($\bar{L}$) where a quantum coin is put in the state:

(3)   Coin State: $\frac{1}{\sqrt{3}}\left(|h\rangle + \sqrt{2}\,|t\rangle\right)$   which is called **C**

$\bar{F}$ "tosses" the coin, and sets a spin ½ particle in the following by the following rules:
   If heads, set to $|\downarrow\rangle$. If tails, set to $|\rightarrow\rangle$. The spin particle is called **S**
The state of the entire system before the experiment begins is written:

$$|\psi_1\rangle = \frac{1}{\sqrt{3}}\left(|h\rangle + \sqrt{2}\,|t\rangle\right)|\bar{F}\rangle\,|S\rangle\,|F\rangle\,|\bar{W}\rangle\,|W\rangle$$

where $F$, $\bar{F}$, $W$ and $\bar{W}$ will be seen to have multiple components. For example, $W$ has two parts, one representing his knowledge of $L$'s state and one representing his knowledge of $\bar{W}$'s state.

For ok/fail type testing in our experiment, we need to define ok/fail bases for the spin states and the coin states respectively:

$$|ok\rangle = \frac{1}{\sqrt{2}}(|\downarrow\rangle - |\uparrow\rangle), \quad |\overline{ok}\rangle = \frac{1}{\sqrt{2}}(|h\rangle - |t\rangle);$$

(*ok* is unbarred as particle is taken to be in unbarred lab

$\overline{ok}$ is barred as coin is taken to be in barred lab)

Note, by picking superposition states, we have a chance of generating our contradiction.

For measuring the state of $\overline{L}$, one must measure the state of the coin *and* $\overline{F}$'s brain. Thus, we have for example:

$$\overline{W} \text{ measures: } |\overline{ok}\rangle_{\overline{L}} = \frac{1}{\sqrt{2}}(|h\rangle|\overline{F}_h\rangle - |t\rangle|\overline{F}_t\rangle),$$

$$\overline{fail}\rangle_{\overline{L}} = |\overline{f}\rangle_{\overline{L}} = \frac{1}{\sqrt{2}}(|h\rangle|\overline{F}_h\rangle + |t\rangle|\overline{F}_t\rangle)$$

For measuring the state of $L$, we must measure the state of the particle's spin *and* $F$'s brain. Thus, we have, for example:

$$W \text{ measures: } \quad |ok\rangle_L = \frac{1}{\sqrt{2}}(|\downarrow\rangle|F_\downarrow\rangle - |\uparrow\rangle|F_\uparrow\rangle)$$

$$|fail\rangle_L \equiv |f\rangle_L = \frac{1}{\sqrt{2}}(|\downarrow\rangle|F_\downarrow\rangle + |\uparrow\rangle|F_\uparrow\rangle)$$

Now, we go through the parts of the experiment as given in box 1 of the paper of DF&RR [1] and show the evolution of the state that corresponds to each step.

Recalling the original state:

(4) $\quad |\psi_1\rangle = \frac{1}{\sqrt{3}}(|h\rangle + \sqrt{2}|t\rangle)|\overline{F}\rangle|S\rangle|F\rangle|\overline{W}\rangle|W\rangle$

The steps are labeled in bold following Box 1 in reference 1.

**(step 00)** $\overline{F}$ measures the coin and sets system *S* per above instructions:

(5) $\quad |\psi_2\rangle = \frac{1}{\sqrt{3}}(|h\rangle|\downarrow\rangle|\overline{F}_h\rangle + \sqrt{2}|t\rangle|\rightarrow\rangle|\overline{F}_t\rangle)|F\rangle|\overline{W}\rangle|W\rangle$

**(step 10)** *F* measure *S*

(6) $\quad |\psi_3\rangle = \frac{1}{\sqrt{3}}(|h\rangle|\downarrow\rangle|\overline{F}_h\rangle|F_\downarrow\rangle + |t\rangle|\downarrow\rangle|\overline{F}_t\rangle|F_\downarrow\rangle + |t\rangle|\uparrow\rangle|\overline{F}_t\rangle|F_\uparrow\rangle)|\overline{W}\rangle|W\rangle$

Decomposing $\psi_3$ into $|\overline{ok}\rangle, |\overline{f}\rangle$, basis for use in next step giving:

$$|\psi_3'\rangle = {}_{\overline{L}}\langle\overline{ok}|\psi_3\rangle|\overline{ok}\rangle_{\overline{L}} + {}_{\overline{L}}\langle\overline{f}|\psi_3\rangle|\overline{f}\rangle_{\overline{L}},$$

(7) $\quad |\psi_3'\rangle = \frac{1}{\sqrt{6}}\left(-|\uparrow\rangle|F_\uparrow\rangle|\overline{ok}\rangle_{\overline{L}} + (2|\downarrow\rangle|F_\downarrow\rangle + |\uparrow\rangle|F_\uparrow\rangle)|\overline{f}\rangle_{\overline{L}}\right)|\overline{W}\rangle|W\rangle$

**(step 20)** $\overline{W}$ measures $\overline{L}$ in the $|\overline{ok}\rangle_{\overline{L}}/|\overline{f}\rangle_{\overline{L}}$ basis:

(8) $\quad |\psi_4\rangle = \frac{1}{\sqrt{6}}\left(-|\uparrow\rangle|F_\uparrow\rangle|\overline{ok}\rangle_{\bar{L}}|\overline{W}_{\overline{ok}}\rangle + (2|\downarrow\rangle\|F_\downarrow\rangle + |\uparrow\rangle|F_\uparrow\rangle)|\bar{f}\rangle_{\bar{L}}|\overline{W}_{\bar{f}}\rangle\right)|W\overline{W}\rangle|W_L\rangle$

Note, here we have split $|W\rangle$ into $|W\overline{W}\rangle|W_L\rangle$ to prepare for the two measurements $W$ will make; one of $\overline{W}$ and one of $L$.

$W$ measures $\overline{W}$ (because $\overline{W}$ announces to $W$) in $|\overline{W}_{\overline{ok}}\rangle/|\overline{W}_{\bar{f}}\rangle$ basis resulting in:

(9)

$|\psi_5\rangle = \frac{1}{\sqrt{6}}\left(-|\uparrow\rangle|F_\uparrow\rangle|\overline{ok}\rangle_{\bar{L}}|\overline{W}_{\overline{ok}}\rangle|W\overline{W}_{\overline{ok}}\rangle + (2|\downarrow\rangle\|F_\downarrow\rangle + |\uparrow\rangle|F_\uparrow\rangle)|\bar{f}\rangle_{\bar{L}}|\overline{W}_{\bar{f}}\rangle|W\overline{W}_{\bar{f}}\rangle\right)|W_L\rangle$

Here: $|W\overline{W}_{\overline{ok}}\rangle$ ($|W\overline{W}_{\bar{f}}\rangle$) represents the state of $W$'s brain associated with him knowing that $\overline{W}$ measures $\bar{L}$ in the fail (ok) state.

Then, ***transforming*** to the $|ok\rangle_L/|f\rangle_L$ basis for use in next step:

$|\psi'_5\rangle = {}_L\langle ok|\psi_5\rangle|ok\rangle_L + {}_L\langle f|\psi_5\rangle|f\rangle_L$

$|ok\rangle_L = \frac{1}{\sqrt{2}}(|\downarrow\rangle|F_\downarrow\rangle - |\uparrow\rangle|F_\uparrow\rangle), \quad |f\rangle_L = \frac{1}{\sqrt{2}}(|\downarrow\rangle|F_\downarrow\rangle + |\uparrow\rangle|F_\uparrow\rangle)$

(10)

$|\psi'_5\rangle = \frac{1}{\sqrt{12}}\begin{pmatrix}|ok\rangle_L|\overline{ok}\rangle_{\bar{L}}|\overline{W}_{\overline{ok}}\rangle|W\overline{W}_{\overline{ok}}\rangle + |ok\rangle_L|\bar{f}\rangle_{\bar{L}}|\overline{W}_{\bar{f}}\rangle|W\overline{W}_{\bar{f}}\rangle - |f\rangle_L|\overline{ok}\rangle_{\bar{L}}|\overline{W}_{\overline{ok}}\rangle|W\overline{W}_{\overline{ok}}\rangle \\ + 3|f\rangle_L|\bar{f}\rangle_{\bar{L}}|\overline{W}_{\bar{f}}\rangle|W\overline{W}_{\bar{f}}\rangle\end{pmatrix}|W_L\rangle$

**(step 30)** $W$ measures $L$ in $|ok\rangle_L/|f\rangle_L$ basis

(11) $\quad |\psi_6\rangle = \frac{1}{\sqrt{12}}\begin{pmatrix}|ok\rangle_L|W_{ok}\rangle|\overline{ok}\rangle_{\bar{L}}|\overline{W}_{\overline{ok}}\rangle|W\overline{W}_{\overline{ok}}\rangle + |ok\rangle_L|W_{ok}\rangle|\bar{f}\rangle_{\bar{L}}|\overline{W}_{\bar{f}}\rangle|W\overline{W}_{\bar{f}}\rangle \\ -|f\rangle_L|W_f\rangle|\overline{ok}\rangle_{\bar{L}}|\overline{W}_{\overline{ok}}\rangle|W\overline{W}_{\overline{ok}}\rangle + 3|f\rangle_L|W_f\rangle|\bar{f}\rangle_{\bar{L}}|\overline{W}_{\bar{f}}\rangle|W\overline{W}_{\bar{f}}\rangle\end{pmatrix}$

$\overline{F}$ measures $W$ ($W$ announces to $\overline{F}$) in $|W_{ok}\rangle/|W_f\rangle$ basis

(Note: up to this point, to simplify notation we have not carried a second ket for $\overline{F}$'s measurement of $W$)

(12) $\quad |\psi_7\rangle = \frac{1}{\sqrt{12}}\begin{pmatrix}\left(|\overline{ok}\rangle_{\bar{L}}|\overline{W}_{\overline{ok}}\rangle|W\overline{W}_{\overline{ok}}\rangle + |\bar{f}\rangle_{\bar{L}}|\overline{W}_{\bar{f}}\rangle|W\overline{W}_{\bar{f}}\rangle\right)|ok\rangle_L|W_{ok}\rangle|\overline{F}W_{ok}\rangle \\ + \left(-|\overline{ok}\rangle_{\bar{L}}|\overline{W}_{\overline{ok}}\rangle|W\overline{W}_{\overline{ok}}\rangle + 3|\bar{f}\rangle_{\bar{L}}|\overline{W}_{\bar{f}}\rangle|W\overline{W}_{\bar{f}}\rangle\right)|f\rangle_L|W_f\rangle|\overline{F}W_f\rangle\end{pmatrix}$

Here, the physical state of $\overline{F}$'s brain associated with him knowing $W$ is in the fail state is written $|\overline{F}W_f\rangle$. The state when he knows $W$ is in the ok state is written: $|\overline{F}W_{ok}\rangle$.

Converting to $|h\rangle|\overline{F}\rangle$ and $|t\rangle|\overline{F}\rangle$ basis:

(13) $|\psi_7'\rangle =$

$$\frac{1}{\sqrt{24}}\left(\begin{pmatrix}|h\rangle|\bar{F}_h\rangle\begin{pmatrix}|ok\rangle_L|W_{ok}\rangle|\bar{F}W_{ok}\rangle\left(|\bar{W}_{\overline{ok}}\rangle|W\bar{W}_{\overline{ok}}\rangle+|\bar{W}_{\bar{f}}\rangle|W\bar{W}_{\bar{f}}\rangle\right)\\+|f\rangle_L|W_f\rangle|\bar{F}W_f\rangle\left(-|\bar{W}_{\overline{ok}}\rangle|W\bar{W}_{\overline{ok}}\rangle+3|\bar{W}_{\bar{f}}\rangle|W\bar{W}_{\bar{f}}\rangle\right)\end{pmatrix}+\\|t\rangle|\bar{F}_t\rangle\begin{pmatrix}|ok\rangle_L|W_{ok}\rangle|\bar{F}W_{ok}\rangle\left(|\bar{W}_{\bar{f}}\rangle|W\bar{W}_{\bar{f}}\rangle-|\bar{W}_{\overline{ok}}\rangle|W\bar{W}_{\overline{ok}}\rangle\right)\\+|f\rangle_L|W_f\rangle|\bar{F}W_f\rangle\left(3|\bar{W}_{\bar{f}}\rangle|W\bar{W}_{\bar{f}}\rangle+|\bar{W}_{\overline{ok}}\rangle|W\bar{W}_{\overline{ok}}\rangle\right)\end{pmatrix}\right)$$

## The EWF State Predictions II
(with $\bar{F}$ collapse of coin state)

Now, we consider the case in which $\bar{F}$'s observation of the coin state collapses that state to an eigenstate (heads or tails). In following the below analysis, recall in our analysis that only he who actually does a measurement collapses the state. Those who merely speculate about someone else's measurement do not collapse the state. In our EWF II analysis below, only $\bar{F}$ collapses a state, so all of the rest of the measurements evolve according to the Schrödinger equation, i.e., with no collapse.

Recall the following notation: to distinguish the below (incorrect) wave functions that invoke collapse to the tails eigenstate from the above (correct) wave functions, we insert a subscript "c" before the number that indexes the stage of the evolution of the system.

The **initial state** remains:

(14) $\quad |\psi_1\rangle = \frac{1}{\sqrt{3}}\left(|h\rangle+\sqrt{2}|t\rangle\right)|\bar{F}\rangle|S\rangle|F\rangle|\bar{W}\rangle|W\rangle$

**(step 00)** $\bar{F}$ measures the coin to be tails and then sets system $S$ to $|\rightarrow\rangle$ per above instructions:

(15) $\quad |\psi_{c2}\rangle = |t\rangle|\rightarrow\rangle|\bar{F}_t\rangle|F\rangle|\bar{W}\rangle|W\rangle = \frac{1}{\sqrt{2}}\left(|\uparrow\rangle+|\downarrow\rangle\right)|t\rangle|\bar{F}_t\rangle|F\rangle|\bar{W}\rangle|W\rangle$

Notice $\bar{F}$ has collapsed the coin state to the tails eigenstate.

**(step 10)** $F$ measure $S$ (assuming no collapse to $+\frac{1}{2}$, i.e., up)
Now, $F$ doesn't deduce anything about his own knowledge, so doesn't apply collapse. No one else deduces anything from $F$'s knowledge so the spin state is not collapsed for anyone.

(16) $\quad |\psi_{c3}\rangle = \frac{1}{\sqrt{2}}\left(|\downarrow\rangle|F_\downarrow\rangle+|\uparrow\rangle|F_\uparrow\rangle\right)|t\rangle|\bar{F}_t\rangle|\bar{W}\rangle|W\rangle$

Decomposing $\psi_3$ into $|\overline{ok}\rangle, |\overline{f}\rangle$, basis gives:

$$|\psi'_{c3}\rangle = \frac{1}{2}\begin{pmatrix}\left(-|\downarrow\rangle|F_\downarrow\rangle-|\uparrow\rangle|F_\uparrow\rangle\right)|\overline{ok}\rangle_{\bar{L}} + \\ \left(+|\downarrow\rangle|F_\downarrow\rangle+|\uparrow\rangle|F_\uparrow\rangle\right)|\bar{f}\rangle_{\bar{L}}\end{pmatrix}|\overline{W}\rangle|W\rangle$$

$$= \frac{1}{2}\left(\left(-|\overline{ok}\rangle_{\bar{L}}+|\bar{f}\rangle_{\bar{L}}\right)\left(|\downarrow\rangle|F_\downarrow\rangle+|\uparrow\rangle|F_\uparrow\rangle\right)\right)|\overline{W}\rangle|W\rangle$$

**(step 20)** $\overline{W}$ measures $\bar{L}$ in the $|\overline{ok}\rangle_{\bar{L}}/|\bar{f}\rangle_{\bar{L}}$ basis:

(17) $\quad |\psi_{c4}\rangle = \frac{1}{2}\left(|\downarrow\rangle|F_\downarrow\rangle+|\uparrow\rangle|F_\uparrow\rangle\right)\left(-|\overline{ok}\rangle_{\bar{L}}|\overline{W}_{\overline{ok}}\rangle+|\bar{f}\rangle_{\bar{L}}|\overline{W}_{\bar{f}}\rangle\right)|W\overline{W}\rangle|W_L\rangle$

(As an aside notice, at this point, due to entanglement with $\overline{W}$, we no longer have: $|\overline{F}_t\rangle|t\rangle = -|\overline{ok}\rangle_{\bar{L}}+|\bar{f}\rangle_{\bar{L}}$ of $\psi_{c3}'$. This reintroduces the heads state, we no longer have a simple tails state!)

$W$ measures $\overline{W}$ (because $\overline{W}$ announces to $W$) in $|\overline{W}_{\overline{ok}}\rangle/|\overline{W}_{\bar{f}}\rangle$ basis resulting in:

(18) $\quad |\psi_{c5}\rangle = \frac{1}{2}\left(|\downarrow\rangle|F_\downarrow\rangle+|\uparrow\rangle|F_\uparrow\rangle\right)\left(-|\overline{ok}\rangle_{\bar{L}}|\overline{W}_{\overline{ok}}\rangle|W\overline{W}_{\overline{ok}}\rangle+|\bar{f}\rangle_{\bar{L}}|\overline{W}_{\bar{f}}\rangle|W\overline{W}_{\bar{f}}\rangle\right)|W_L\rangle$

Then, transforming to the $|ok\rangle_L/|f\rangle_L$ basis for use in next step:

$$|\psi'_5\rangle = {}_L\langle ok|\psi_5\rangle|ok\rangle_L + {}_L\langle f|\psi_5\rangle|f\rangle_L$$

$$|ok\rangle_L = \frac{1}{\sqrt{2}}\left(|\downarrow\rangle|F_\downarrow\rangle-|\uparrow\rangle|F_\uparrow\rangle\right), |f\rangle_L = \frac{1}{\sqrt{2}}\left(|\downarrow\rangle|F_\downarrow\rangle+|\uparrow\rangle|F_\uparrow\rangle\right)$$

(19) $\quad |\psi'_{c5}\rangle = \frac{1}{\sqrt{2}}\left(-|\overline{ok}\rangle_{\bar{L}}|\overline{W}_{\overline{ok}}\rangle|W\overline{W}_{\overline{ok}}\rangle+|\bar{f}\rangle_{\bar{L}}|\overline{W}_{\bar{f}}\rangle|W\overline{W}_{\bar{f}}\rangle\right)|f\rangle_L|W_L\rangle$

**(step 30)** $W$ measures $L$ in $|ok\rangle_L/|f\rangle_L$ basis:

(20) $\quad |\psi_{c6}\rangle = \frac{1}{\sqrt{2}}\left(-|\overline{ok}\rangle_{\bar{L}}|\overline{W}_{\overline{ok}}\rangle|W\overline{W}_{\overline{ok}}\rangle+|\bar{f}\rangle_{\bar{L}}|\overline{W}_{\bar{f}}\rangle|W\overline{W}_{\bar{f}}\rangle\right)|f\rangle_L|W_f\rangle$

$\bar{F}$ measures $W$ ($W$ announces to $\bar{F}$) in $|W_{ok}\rangle/|W_f\rangle$ basis

(For $\bar{F}$, the state will collapse to the eigenstate corresponding to the value he measures, but since we do not know what value he gets, we keep it pre-collapsed. NB, this is the last state that this system enters and thus collapsed state is not needed for input to another calculation.)

The physical state of $\bar{F}$'s brain associated with him knowing $W$ is in the fail state is written $|\bar{F}W_f\rangle$. The state when he knows $W$ is in the ok state is written: $|\bar{F}W_{ok}\rangle$.

(21) $\quad |\psi_{c7}\rangle = \frac{1}{\sqrt{2}}\left(-|\overline{ok}\rangle_{\bar{L}}|\overline{W}_{\overline{ok}}\rangle|W\overline{W}_{\overline{ok}}\rangle+|\bar{f}\rangle_{\bar{L}}|\overline{W}_{\bar{f}}\rangle|W\overline{W}_{\bar{f}}\rangle\right)|f\rangle_L|W_f\rangle|\bar{F}W_f\rangle$

Convert to $|h\rangle|\bar{F}\rangle$ and $|t\rangle|\bar{F}\rangle$ basis using

$$\left|\overline{ok}\right\rangle_{\overline{L}} = \frac{1}{\sqrt{2}}\left(|h\rangle|\overline{F}_h\rangle - |t\rangle|\overline{F}_t\rangle\right), |\overline{f}\rangle_{\overline{L}} = \frac{1}{\sqrt{2}}\left(|h\rangle\|\overline{F}_h\rangle + |t\rangle|\overline{F}_t\rangle\right)$$

$$|\psi'_{c7}\rangle = \frac{1}{2}\left(-\left(|h\rangle|\overline{F}_h\rangle - |t\rangle|\overline{F}_t\rangle\right)|\overline{W}_{\overline{ok}}\rangle|W\overline{W}_{\overline{ok}}\rangle + \left(|h\rangle|\overline{F}_h\rangle + |t\rangle|\overline{F}_t\rangle\right)|\overline{W}_{\overline{f}}\rangle|W\overline{W}_{\overline{f}}\rangle\right)|f\rangle_L|W_f\rangle|\overline{F}W_f\rangle$$

(22) $\quad |\psi'_{c7}\rangle = \frac{1}{2}\begin{pmatrix}|h\rangle|\overline{F}_h\rangle\left(-|\overline{W}_{\overline{ok}}\rangle|W\overline{W}_{\overline{ok}}\rangle + |\overline{W}_{\overline{f}}\rangle|W\overline{W}_{\overline{f}}\rangle\right) \\ +|t\rangle|\overline{F}_t\rangle\left(|\overline{W}_{\overline{ok}}\rangle|W\overline{W}_{\overline{ok}}\rangle + \overline{W}_{\overline{f}}\rangle|W\overline{W}_{\overline{f}}\rangle\right)\end{pmatrix}|f\rangle_L|W_f\rangle|\overline{F}W_f\rangle$

## Conclusion

We have seen the evolution of the states with and without collapse and have seen that the ad hoc nature of the collapse postulate creates a confusion and finally a contradiction in QM, at least in the way that DF&RR apply it. By contrast, a straightforward statistical interpretation, the ensemble interpretation, gives no contradiction and gives a clear understanding of what happens. As explained in a previous paper,[3] the SE development of the wave function has no collapse, so there is no reason to insert one. In that sense, inserting one actually constitutes introducing a new theory, one distinct from ordinary quantum mechanics.

The contradictions that result from the no-go theorem should be a cautionary tale against the use of the idea of collapse in foundational discussions of QM. It can serve to remind us that our habits lean heavily towards a collapse view, even when we (e.g., DF&RR) are discussing interpretations such as the many worlds or de Broglie Bohm.[12] Working through the theorem in light of the ensemble interpretation can help us amend those habits and see how much simpler QM can actually be without the ad hoc collapse postulate. QM is obviously a statistical theory, measurement is an integral part of QM and is, thus, naturally also treated statistically. Moreover, students can follow the argument and its resolution, and thus can also learn these things, as well as how to properly derive, in some detail, the state evolution that occurs during measurements.

**Acknowledgements:**
I would like to acknowledge Doyl Dickel for very helpful conversations during the writing of this article and for his insight-generating comments on the completed article.

---

[12] DF&RR discuss, among other interpretations: de Broglie Bohm, many worlds, consistent histories, QBism and ETH approach.